\documentclass{cs19proc}

\usepackage{kantlipsum}

\editors{G.~A. Feiden}
\publisher{Zenodo}
\conference{The 19th Cambridge Workshop on Cool Stars, Stellar Systems, and the Sun}
\conferencedate{2016}

\title{Statistical properties of superflares on solar-type stars with Kepler data}
\author{Yuta Notsu$^{1, 2, *}$, 
Hiroyuki Maehara$^{3}$, 
Takuya Shibayama$^{2, 4}$, 
Satoshi Honda$^{5}$, 
Shota Notsu$^{1, 2}$, 
Kosuke Namekata$^{1}$, 
Daisaku Nogami$^{1}$, 
Kazunari Shibata$^{6}$  
}

\affiliation{$^{1}$ Department of Astronomy, Kyoto University, Kitashirakawa-Oiwake-cho, Sakyo-ku, Kyoto 606-8502, Japan \\
$^{2}$ JSPS Research Fellow (DC1) \\
$^{3}$ Okayama Astrophysical Observatory, National Astronomical Observatory of Japan, 3037-5 Honjo, Kamogata, Asakuchi, Okayama 719-0232, Japan \\
$^{4}$ Institute for Space-Earth Environmental Research, Nagoya University, Furo-cho, Chikusa-ku, Nagoya, Aichi, Japan, 464-8601, Japan \\
$^{5}$ Center for Astronomy, University of Hyogo, 407-2, Nishigaichi, Sayo-cho, Sayo, Hyogo 679-5313, Japan \\
$^{6}$ Kwasan and Hida Observatories, Kyoto University, Yamashina-ku, Kyoto 607-8471, Japan \\ 
$^{*}$ ynotsu@kwasan.kyoto-u.ac.jp \\
}

\shorttitle{Superflares on solar-type stars}
\shortauthors{Yuta Notsu et al.}

\abs{
\ \ \ \ \
Superflares are flares that release total energy 10$\sim$10$^{4}$ times greater than
that of the biggest solar flares with energy of $\sim$10$^{32}$ erg.
We searched superflares on solar-type stars (G-type main sequence stars) 
using the Kepler 30-min (long) and 1-min (short) cadence data. 
We found more than 1500 superflares on 279 stars from 30-min cadence data (Q0-6)
and 187 superflares on 23 stars from 1-min cadence data (Q0-17). 
The bolometric energy of detected superflares ranges from the order of 10$^{32}$ erg to 10$^{36}$ erg. 
Using these data, we found that the occurrence frequency ($dN/dE$) of superflares is
expressed as a power-law function of flare energy ($E$) with the index of -1.5 for $10^{33}<E<10^{36}$ erg. 
Most of the superflare stars show quasi-periodic light variations with the amplitude of a few percent,
which can be explained by the rotation of the star with large starspots. 
\\ 
\ \ \ \ \
The bolometric energy released by flares is consistent with the magnetic energy stored around such large starspots.
Furthermore, our analyses indicate that the occurrence frequency of superflares depends on the rotation period,
and that the flare frequency increases as the rotation period decreases. 
However, the energy of the largest flares observed in a given period bin does not show any clear correlation with the rotation period.
We also found that the duration of superflares increases with the flare energy as $E^{0.39+/-0.03}$. 
This can be explained if we assume the time-scale of flares is determined by the Alfv$\acute{\rm{e}}$n time.
}

\begin{document}

\maketitle

\section{Introduction}\label{sec:intro}
Solar flares are energetic explosions in the solar atmosphere 
and are thought to occur by impulsive releases of magnetic energy stored around sunspots (e.g., \citealt{Priest1981}; \citealt{Shibata2011}; \citealt{Shibata2016}).
Typical total energy released by solar flares are $10^{29}\sim10^{32}$ erg.
The occurrence frequency of solar flares decreases as the flare energy increases. 
The frequency energy distribution of solar flares can be fitted by a simple power-law function ($dN/dE\propto E^{-\alpha}$)
with an index ($\alpha$) of $-1.5\sim-1.9$ in the flare energy range between 10$^{24}$ and 10$^{32}$ erg 
(e.g., \citealt{Aschwanden2000}; \citealt{Shimizu1995}; \citealt{Crosby1993}).
The total bolometric energy released by the largest solar flares 
is estimated to be the order of 10$^{32}$ erg \citep{Emslie2012},
and the occurrence frequency of such flares is about once in 10 years.
\\ 
\ \ \ \ \
Stars other than the Sun show flares similar to solar flares (e.g., \citealt{Gershberg2005}).
Many young stars, 
active M-dwarfs known as flare stars,
and binary stars such as RS CVn-type binary stars have ``superflares".
Superflares are flares that have a total energy of $10^{33}\sim10^{38}$ erg \citep{Schaefer2000}, 
10-$10^6$ times larger than that of the largest solar flares on the Sun ($\sim10^{32}$erg; \citealt{Emslie2012}).
Such stars generally rotate fast ($P_{\rm{rot}}\sim$a few days and $v \sin i\gtrsim$10km s$^{-1}$)
and magnetic fields of a few kG are considered to be distributed in large regions on the stellar surface (\citealt{Gershberg2005}).
In contrast, the Sun slowly rotates ($P_{\rm{rot}}\sim$25 days and $v \sin i\sim$2km s$^{-1}$), 
and the mean magnetic field is weak (a few G). 
Then it had been thought that superflares cannot occur on slowly-rotating G-type main-sequence stars like the Sun.
\\ 
\ \ \ \ \ 
\citet{Schaefer2000} reported nine superflare candidates on ordinary solar-type stars 
(F8-G8 main-sequence stars).
The data sources of their study were, however, various and somewhat ambiguous
(e.g., photography, X-ray, visual), and the number of discovered
flares is too small to investigate statistical properties, such as the frequency and energy distribution of superflares.
Considering these previous researches, we have analyzed the lightcurve data of solar-type
stars obtained with the Kepler space telescope. 
We have aimed to find superflares on ordinary solar-type stars, and to investigate 
whether such superflares can occur on the Sun from the statistical points of view.
In the following, we summarized these results, which we have reported on the basis of our analyses of Kepler photometric data 
(cf., \citealt{Maehara2012} \& \citeyear{Maehara2015}; \citealt{Shibata2013}; \citealt{Shibayama2013}; \citealt{YNotsu2013}).

\begin{figure*}[htbp]
 \centering
\includegraphics[width=0.47\linewidth]{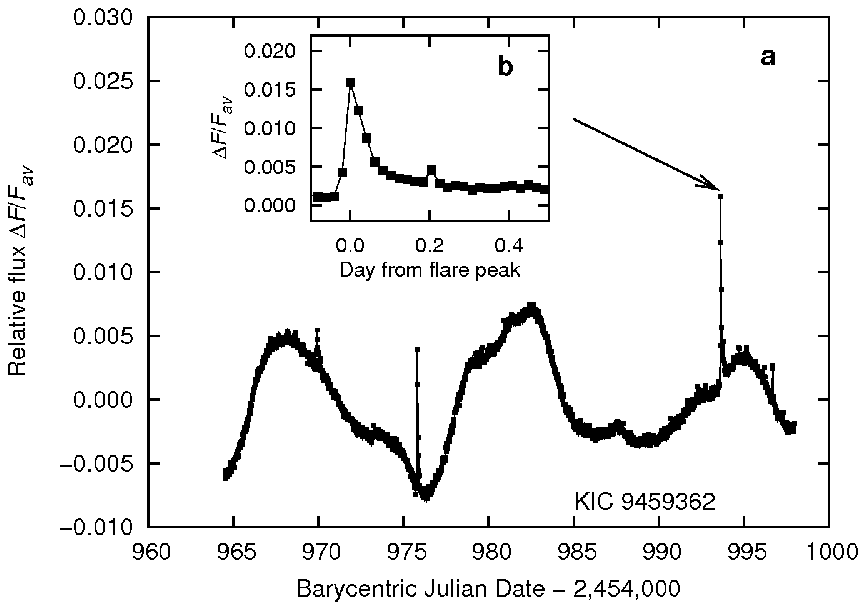} 
\includegraphics[width=0.47\linewidth]{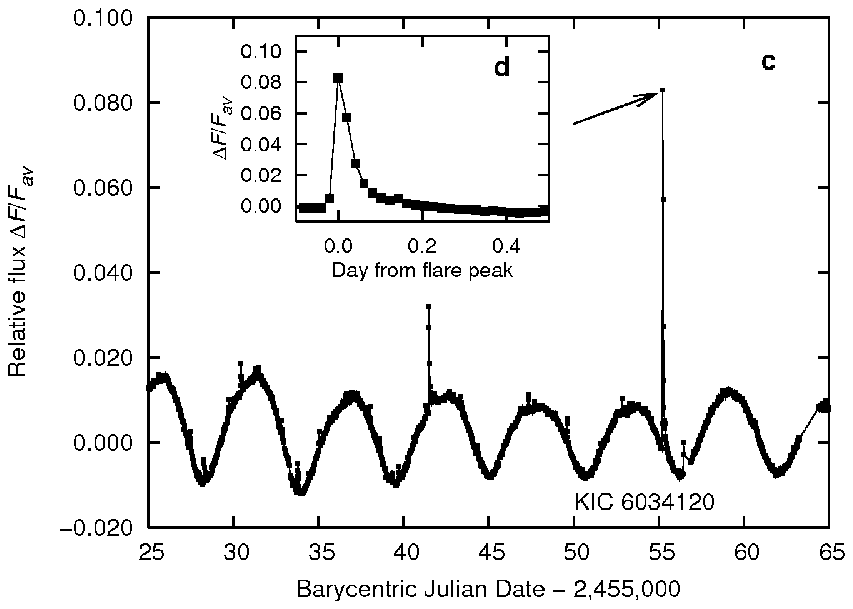} 
\caption{Lightcurves of typical superflares (\citealt{Maehara2012}). 
(a)\&(c): Long-term brightness variations of the solar-type superflare stars (a)  KIC9459362 and (c) KIC6034120. 
(b)\&(d): Enlarged lightcurve of a superflare on (b) KIC9459362 and that on (d) KIC6034120 (indicated by the arrow in panel (a) and (c)).}
\label{fig:lc-Maehara2012}
\end{figure*}

\section{Data and discovery of superflares}\label{sec:data}
We first searched for superflares on solar-type (G-type main sequence) stars 
using the long-cadence (30-min time resolution) data of Kepler space telescope observed for the first 500 days (Quarters 0$\sim$6) \citep{Koch2010}.
Kepler is very useful for detecting small increases in the stellar brightness caused by stellar flares, 
because Kepler realized high photometric precision exceeding 0.01\% for moderately bright stars, 
and obtained continuous time-series data of many stars over a long period \citep{Koch2010}.
As a result, we discovered 1547 superflare events on 279 solar-type (G-type main-sequence) stars 
from the data of $\sim$90,000 solar-type stars (\citealt{Maehara2012}; \citealt{Shibayama2013}).
This number is much larger than the 9 flare candidates on solar-type stars discovered by \citet{Schaefer2000}.
We here define ``{\it solar-type stars}" as stars that have a surface temperature of $5100\leq T_{\rm{eff}}\leq 6000$K 
and a surface gravity of $\log g \geq 4.0$ on the basis of Kepler Input Catalog \citep{Brown2011}.
Figure \ref{fig:lc-Maehara2012} shows typical examples of superflares observed by Kepler, 
which show spike-like brightness increases whose amplitude and duration are several percent and several hours, respectively.
It should be noted that even one of the largest solar flares in recent 20 years (28 October 2003 X17 flare) 
showed only 0.03 percent increase in the total solar irradiance (\citealt{Kopp2005}).
The total energy of these superflares in Figure \ref{fig:lc-Maehara2012} are estimated
to be $5.6\times10^{34}$ erg and $3.0\times10^{35}$ erg \citep{Maehara2012}, 
which are 100$\sim$1000 times larger than the largest solar flare ($\sim$10$^{32}$ erg).

\begin{figure*}
	\centering
	\includegraphics[width=0.7\linewidth]{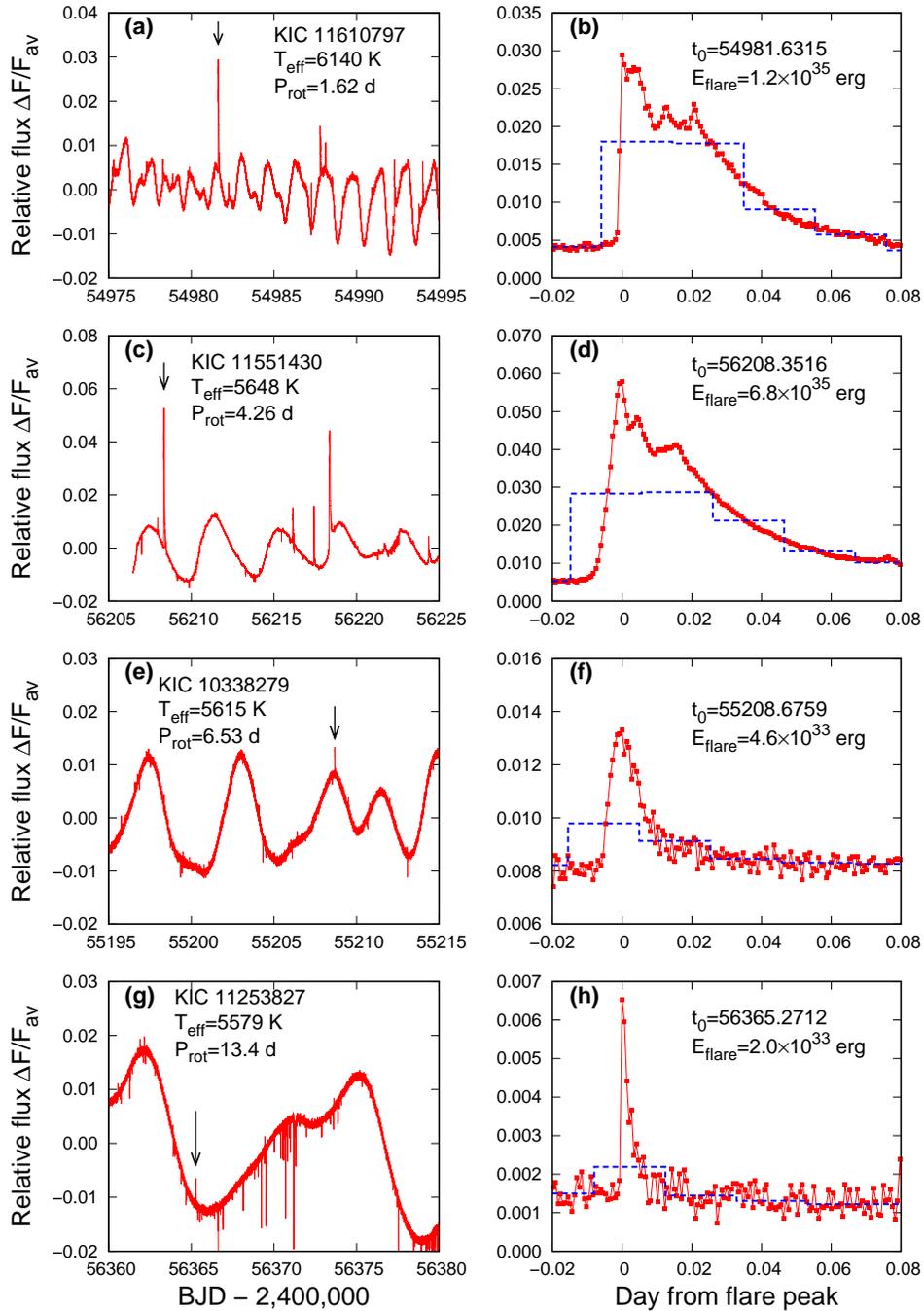}
	\caption{
(a), (c), (e), (g): Long-term light variations of the solar-type superflare stars 
(a)KIC 11610797, (c) KIC 11551430, (e) KIC 10338279, and (g) KIC 11253827 \citep{Maehara2015} . 
(b), (d), (f), (h): Enlarged light curve of a superflare on (b) KIC 11610797, (d) KIC11551430, (f) KIC 10338279, and (h) KIC 11253827 
(indicated by the down arrow in panel (a), (c), (e), and (g)). 
Red filled-squares with solid-lines and blue dashed-lines represent the light curves from short- and long-cadence data, respectively.
	}\label{fig:poster-fig3}
\end{figure*}

\ \ \ \ \ 
In the above analyses (\citealt{Maehara2012}; \citealt{Shibayama2013}), 
we only used the Kepler {\it long} time cadence data with 30-min sampling, 
and it was difficult to detect superflares with energy between 10$^{32}$ to 10$^{34}$ erg due to low time resolution of the data.
Then in \citet{Maehara2015}, we newly analyzed the Kepler {\it short} time cadence data with 1-min sampling \citep{Gilliland2010}.
We found 187 superflares on 23 solar-type stars from the data of 1547 solar-type stars.
Figure \ref{fig:poster-fig3} shows typical light curves of superflares detected from short-cadence data. 
The amplitude of flares normalized by the average brightness of the star range from 
$1.3\times10^{-3}$ to $8.5\times10^{-2}$ and the bolometric energies of flares 
range from $2\times10^{32}$ to $8\times10^{35}$ erg.
The lower end of the amplitude of detected flare is comparable to that of the X17 class 
solar flare (normalized amplitude $\sim 3 \times 10^{-4}$) occurred on Oct 28, 2003 \citep{Kopp2005}.
As shown in Figures \ref{fig:poster-fig3} (b) \& (d), some flares show multiple peaks with 
the peak separation of 100 - 1000 seconds.
On the analogy of the solar flare oscillations, 
these multiple peaks may be caused by coronal oscillations associated with flares (cf. \citealt{Balona2015}; \citealt{Pugh2016}). 

\section{Results and Discussion}\label{sec:results-discussion}
\subsection{Brightness variation of superflare stars}\label{subsec:lightcurve}
\begin{figure}[htbp]
 \begin{center}
\includegraphics[width=0.5\linewidth]{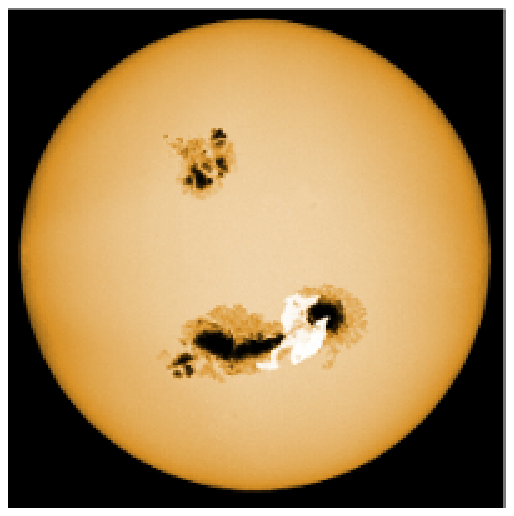} 
 \end{center}
\caption{
Artistic view of a superflare and big starspots on a solar-type star
on the basis of Kepler observations (courtesy of Hiroyuki Maehara).
}\label{fig:flare-img}
\end{figure}

We then performed more analyses in order to investigate whether properties of
superflare stars can be explained by applying our current physical understanding of the Sun. 
Solar flares are the intense releases of magnetic energy stored around sunspots, as mentioned above. 
If we assume superflares are the similar magnetic energy releases, 
large starspots are necessary to explain their large amount of energy.
Many of superflare stars show quasi-periodic brightness variations 
with a typical period of from one day to a few tens of days, as seen in Figures \ref{fig:lc-Maehara2012} \& \ref{fig:poster-fig3}.
The amplitude of these brightness variations is in the range of 0.1-10\% \citep{Maehara2012}, 
and is much more larger than that of solar brightness variation (0.01-0.1\%; e.g., \citet{Lanza2003}) 
caused by the existence of sunspots on the rotating solar surface.
Such stellar brightness variations can be explained by the rotation of a star with large starspots (Figure \ref{fig:flare-img}).
Then in \citet{YNotsu2013}, we developed this idea by performing simple calculations of brightness variation of the rotating star with large starspots,
and showed that these brightness variations of superflare stars can be well explained by the rotation of a star with fairly large starspots, 
taking into account the effects of the inclination angle and the spot latitude.

These interpretations described here are now supported by the spectroscopic studies.
We observed 50 solar-type superflare stars 
by using Subaru/HDS (\citealt{SNotsu2013}; \citealt{Nogami2014}; \citealt{Notsu2015a} \& \citeyear{Notsu2015b}; \citealt{Honda2015}). 
We first found that more than half of the targets have no evidence of binary system, 
and stellar atmospheric parameters (temperature, surface gravity, and metallicity) of these stars 
are in the range of ordinary solar-type stars \citep{Notsu2015a}. 
Then \citet{Notsu2015b} supported the above interpretation that 
the quasi-periodic brightness variations of superflare stars are explained by the rotation of a star with large starspots,
by measuring $v\sin i$ (projected rotational velocity) and the intensity of Ca II 8542 line.
Existence of large starspots on superflare stars were also supported by \citet{Karoff2016} using Ca II H\& K observations with LAMOST telescope. 
Then in the following, we use rotation period and starspot coverage of each star estimated from the above brightness variations, 
and discuss statistical properties of superflares and superflare-generating stars.

\subsection{Relation between flare energy and area of starspots}\label{subsec:fene}
\begin{figure}
	\centering
	\includegraphics[width=0.99\linewidth]{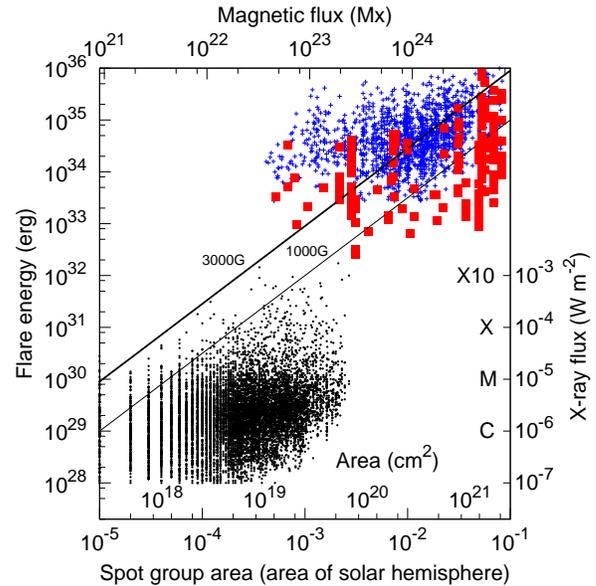}
	\caption{
Scatter plot of flare energy as a function of spot area (\citealt{Maehara2015}; \citealt{YNotsu2013}).
The lower and upper horizontal axis indicate the area of starspot group in the unit of the area 
of solar hemisphere and the magnetic flux for $B=3000$ G, respectively.
The vertical axis represents the bolometric energy released by each flare.
Red filled-squares and blue small-crosses indicate superflares on solar-type stars
detected from short-cadence \citep{Maehara2015} and long-cadence data \citep{Shibayama2013}, respectively.
Small black filled-circles represent solar flares (Ishii et al. private communication, 
based on the data retrieved from the website of NOAA/NGDC, 
Solar-Terrestrial Physics Division at $\langle$http://www.ngdc.noaa.gov/stp/$\rangle$).
The area of starspot group for G-type main sequence stars was estimated from the full amplitude of brightness variations
during each observation period in which the flare occurred.
We assumed that bolometric energies of B, C, M, X, and X10 class solar flares 
are $10^{28}$, $10^{29}$, $10^{30}$, $10^{31}$, and $10^{32}$ erg
from observational estimates of energies of typical solar flares (\citealt{Benz2008}; \citealt{Emslie2012}).
	}
	\label{fig:poster-fig8}
\end{figure}

\ \ \ \ \ \ \
Figure \ref{fig:poster-fig8} shows the scatter plot of flare energy of solar flares and superflares as a function of the area of the sunspot and starspot group ($A_{\rm{spot}}$).  
The area of starspots on superflare stars were estimated from the amplitude of brightness variations.
Although there is a large scatter, the majority of energetic flares occur on the stars with large starspots.
Since the flares are a result of a sudden conversion of magnetic energy to other forms (i.e., thermal and kinetic energy; cf. \citealt{Shibata2011}),
the bolometric energy released by flares may be a fraction of the magnetic energy ($E_{\rm{mag}}$) (\citealt{Shibata2013}; \citealt{YNotsu2013}; \citealt{Maehara2015}).

\ \ \ \ \ \ \
Here, the energy released by a flare can be written as:
\begin{eqnarray}\label{eq:Eflare-1}
E_{\rm{flare}} \sim fE_{\rm{mag}} \sim \frac{fB^{2}L^{3}}{8\pi} \sim \frac{fB^{2}A_{\rm{spot}}^{3/2}}{8\pi} \ , 
\end{eqnarray}
where $f$, $B$, and $L$ correspond to the fraction of magnetic energy released by the flare,
the magnetic field strength of the starspots,
and the scale length of the starspot group, respectively (\citealt{Shibata2013}; \citealt{YNotsu2013}). 
We here also roughly assume $L^{2} \sim A_{\rm{spot}}$(the total area of the starspot group).
The Equation (\ref{eq:Eflare-1}) suggests that the flare energy is roughly proportional
to the area of the starspot group to the power of 3/2.
Using the typical values, Equation (\ref{eq:Eflare-1}) can be written as:
\begin{eqnarray}\label{eq:Eflare-2}
E_{\rm flare}  &\sim& 7\times 10^{32} {\rm (erg)} \left(\frac{f}{0.1}\right) \nonumber \\ && \times\left(\frac{B}{1000{\rm G}}\right)^2 \left[\frac{A_{\rm spot}/2\pi R_{\odot}^2}{0.001}\right]^{3/2} \ .
\end{eqnarray}
\\
\ \ \ \ \ \ \ 
Solid and dashed lines in Figure \ref{fig:poster-fig8} represent Equation (\ref{eq:Eflare-2}) for $f = 0.1$, $B = 1000 G$, and $f = 0.1$, $B = 3000 G$, respectively. 
Majority of superflares (Red filled-squares and blue small-crosses) 
and almost all solar flares (black small filled-circles) are below these analytic lines. 
This suggests that the upper limit of the energy released by the flare is basically
comparable to the stored magnetic energy estimated from the area of starspots. 
There are, however, some superflares above these analytic lines. 
These superflares would occur on the stars with low-inclination angle or stars with starspots around polar region. 
In the case of such stars, the brightness variation caused by the rotation becomes small. 
Hence the actual area of starspots should be larger than that estimated from the amplitude of brightness variation.
The effect of stellar inclination on Figure \ref{fig:poster-fig8} are confirmed with spectroscopic data in \citet{Notsu2015b}.
\\
\ \ \ \ \ \ \
As a result, Figure \ref{fig:poster-fig8} shows that both solar and stellar flares are caused by the release of 
magnetic energy stored around spots.
This result is consistent with the magnetic reconnection model of stellar flares (e.g., \citealt{Shibata2011}; \citealt{Shibata2016}).

\subsection{Correlation between flare duration and flare energy}\label{subsec:duration-fene}
\begin{figure}
	\centering
	\includegraphics[width=0.95\linewidth]{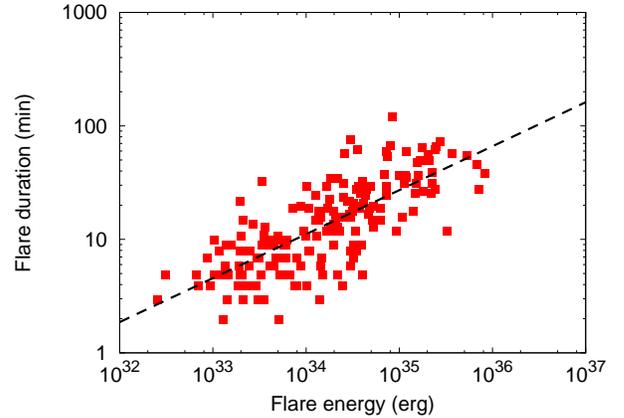}
	\caption{
	Scatter plot of the duration of superflares as a function of the bolometric energy (\citealt{Maehara2015}).
Filled-squares indicate superflares on solar-type stars from short-cadence data. 
We used e-folding decay time as the flare duration.
Dotted line indicates the linear regression for the data of superflares from short-cadence data. 
The power-law slope of the line is $0.39\pm 0.03$.
	}
	\label{fig:poster-fig10}
\end{figure}

Figure \ref{fig:poster-fig10} shows the duration (e-folding decay time) of superflares 
as a function of the bolometric energy of flares, on the basis of Kepler 1-min (short) cadence data \citep{Maehara2015}. 
A linear fit for the superflares from short-cadence data in the log-log plot shows
\begin{eqnarray}\label{eq:tau-flare-fit}
\tau _{\rm{flare}} \propto E^{0.39\pm 0.03}_{\rm{flare}} \ \,
\end{eqnarray}
where $\tau _{\rm{flare}}$ and $E _{\rm{flare}}$ are the duration and energy of superflares. 
Similar correlation between the flare duration and energy was observed in solar flares 
(e.g., \citealt{Veronig2002}, \citealt{Christe2008}) \footnote{
However, there have been no well-known statistical researches discussing the correlation between the flare duration and energy 
with solar ``white light" data, while stellar flares observed by Kepler are all ``white light flares".
This is because it has been difficult to observe solar white-light flares since this requires high spatial and temporal resolution.
\citet{Veronig2002} used soft X-ray fluence data observed with GOES satellite, 
and \citet{Christe2008} used the duration and peak flux of RESSI satellite (hard X-ray). 
We note here that more observations and statistical analyses of solar white-light flares are necessary
for more detailed discussions.
}. These similarity between solar flares and superflares on solar-type stars suggests 
that solar flares and superflares are caused by the same physical process (i.e., reconnection).

\ \ \ \ \ 
The flare energy is related to the magnetic energy stored around the starspots, as shown in Equation (\ref{eq:Eflare-1}).
On the other hand, the duration of flares, especially the duration of impulsive phase of flares, 
is comparable to the reconnection time ($\tau_{\rm{rec}}$) that can be written as:
\begin{eqnarray}\label{eq:tau-rec}
\tau _{\rm flare} \sim \tau _{\rm rec} \sim \tau _{\rm A}/M_{\rm A} \sim L/v_{\rm A}/M_{\rm A} \,
\end{eqnarray}
where $\tau _{\rm A} = L/v_{\rm A}$ is the Alfv$\acute{\rm e}$n time, $v_{\rm A}$ is the Alfv$\acute{\rm e}$n velocity, 
and $M_{\rm A}$ is the non-dimensional reconnection rate ($M_{\rm A}\sim 0.1-0.01$ in the corona \citep{Petschek1964}).
If we assume $B$ and $v_{\rm A}$ are not so different among solar-type stars, 
the duration of flares can be written as
\begin{equation}\label{tau-E-theory}
\tau _{\rm flare} \propto E_{\rm flare} ^{1/3}.
\end{equation}
from Equations (\ref{eq:Eflare-1}) and (\ref{eq:tau-rec}).
This suggests that the power-law slope for the correlation
between the flare duration and flare energy is about $1/3$,
and this is comparable to the observed value of $0.39\pm 0.03$.
\\
\subsection{Occurrence frequency distribution of superflares}\label{subsec:frequency}
\begin{figure}
	\centering
	\includegraphics[width=0.95\linewidth]{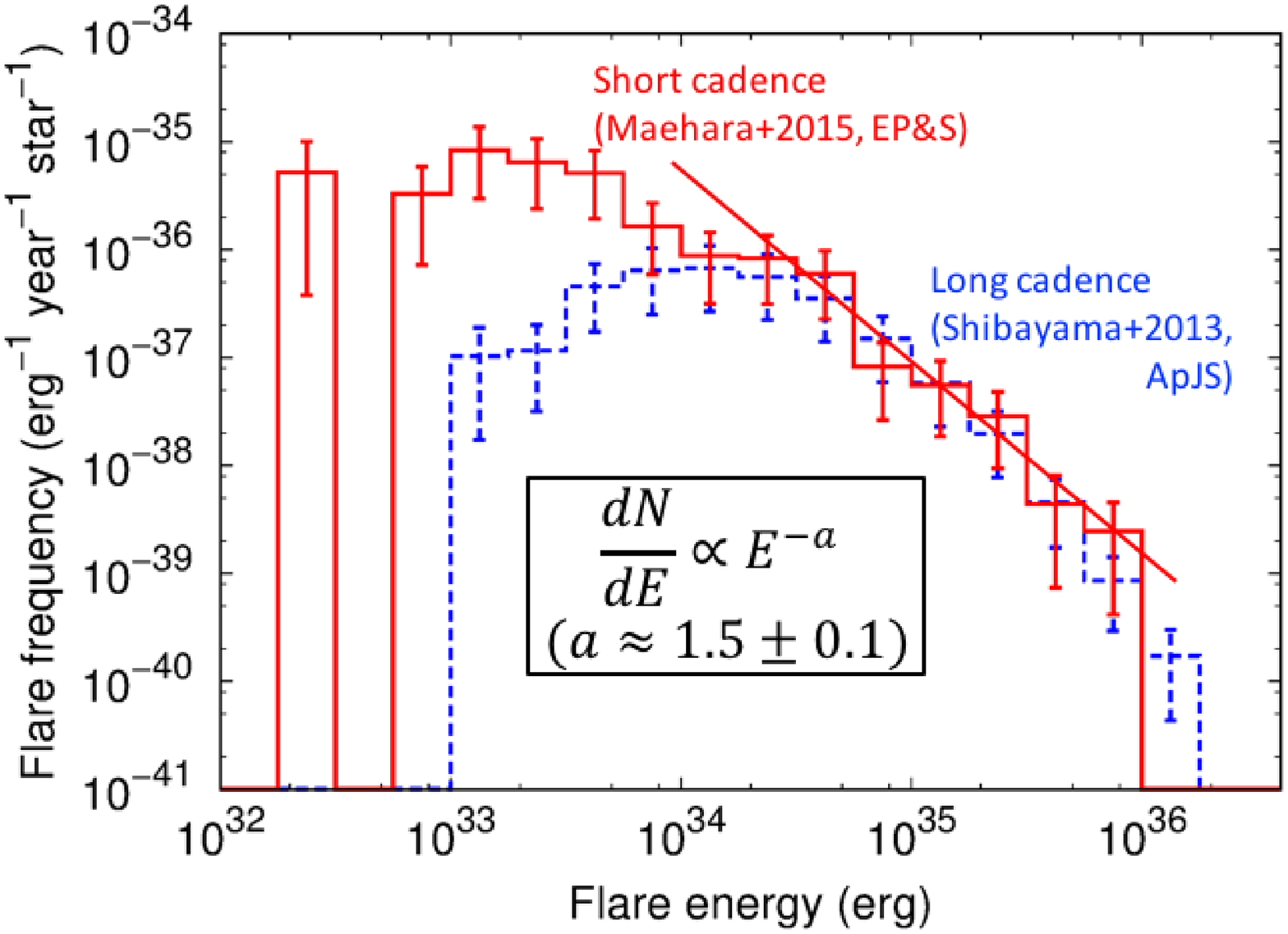}
	\caption{Red solid and blue dashed histograms represent the occurrence frequency of superflares
on all solar-type stars from short- (\citealt{Maehara2015}) 
and long-cadence data (\citealt{Shibayama2013}) as a function of total energy of superflares, respectively. 
The vertical axis indicates the number of superflares per star, per year, and per unit energy. 
Since the period distribution of the stars observed in short-cadence mode is biased to the shorter-period end, 
we estimate the corrected occurrence frequency for short cadence data taking into account the bias 
in the period distribution of the observed samples (For the details, see \citet{Maehara2015}).}
	\label{fig:poster-fig5}
\end{figure}

The analyses of Kepler data enabled us to discuss statistical properties of superflares 
since a large number of flare events were discovered.
Figure \ref{fig:poster-fig5} shows the occurrence frequency of superflares ($dN/dE$) 
as a function of the bolometric energy of superflares ($E$).
Red solid histogram in Figure \ref{fig:poster-fig5} represents the
frequency distribution of superflares on all solar-type stars derived from short-cadence data \citep{Maehara2015} 
and blue dashed histogram represents that from long-cadence data (\citealt{Shibayama2013}).
The frequency of flares derived from the long-cadence data is less than that from
the short-cadence data for the flare energy below $10^{34}$ erg. 
This difference is caused by the detection limit of flare search method. 
We can detect smaller flares more accurately with short-cadence data, which have much better time resolution, 
as seen in Figures \ref{fig:poster-fig3} (f) \& (h).
In Figure \ref{fig:poster-fig5}, both frequency distributions from short- and long-cadence data are almost the same for the flare energy 
between 10$^{34}$ erg and 10$^{36}$ erg, 
and can be fitted by a power-law function ($dN/dE \propto E^{-\alpha}$).
Using the combined data set from both short- and long-cadence data, 
the power-law index $\alpha$ is $1.5\pm0.1$ for the flare energy of $4\times 10^{33} \sim 1\times 10^{36}$ erg.
\\

\begin{figure}
	\centering
	\includegraphics[width=0.95\linewidth]{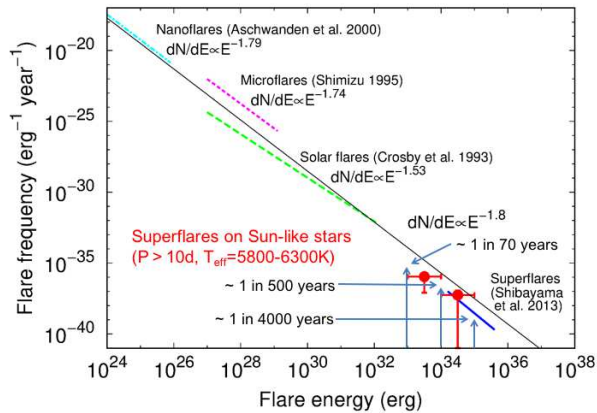}
	\caption{Comparison between occurrence frequency of superflares on Sun-like stars and that of solar flares \citep{Maehara2015}. 
Red filled-circles indicates the occurrence frequency of superflares on Sun-like stars 
(solar-type stars with $P_{\rm{rot}}\geq$10 days and $5800 \leq T_{\rm{eff}} < 6300$K) derived from Kepler short-cadence data. 
Blue bold-solid line represents the power-law frequency distribution of superflares on Sun-like stars derived from Kepler long-cadence data \citep{Shibayama2013}.
Dashed lines indicate the power-law frequency distribution of solar flares observed in hard X-ray \citep{Crosby1993}, 
soft X-ray microflares \citep{Shimizu1995}, and EUV nanoflaeres \citep{Aschwanden2000}. 
Occurrence frequency distributions of superflares on Sun-like stars and solar flares are roughly 
on the same power-law line with an index of $-1.8$ (thin-solid line; \citealt{Shibata2013}) 
for the wide energy range between $10^{24}$ erg and $10^{35}$ erg.}
	\label{fig:poster-fig6}
\end{figure}

\ \ \ \ \ 
In Figure \ref{fig:poster-fig6}, 
we compare the frequency distribution of superflares on Sun-like stars (early G-dwarfs with $P_{\rm{rot}}\geq 10$ days) 
derived from both short-cadence (red filled circles) and long-cadence (blue solid line) data with 
solar flares, microflares and nanoflares (dashed lines).
It is remarkable to see that occurrence frequency distributions of superflares on Sun-like stars 
and solar flares are roughly on the same power-law line ($dN/dE\propto E^{-1.8}$; \citealt{Shibata2013}) 
for the wide energy range between $10^{24}$ erg and $10^{35}$ erg.
This distribution suggests that average occurrence frequency of superflares on Sun-like stars with bolometric energy of 
$10^{33}$ erg,  $10^{34}$ erg, and $10^{35}$ erg are once in $\sim 70$ years, $\sim 500$ years, and $\sim 4000$ years, respectively.

\subsection{Dependence of the flare energy and flare frequency on rotation period}\label{sec-Kepler:Prot-flare}
Previous observations of solar-type stars (e.g., \citealt{Pallavicini1981}; \citealt{Pizzolato2003}; \citealt{Wright2011}) 
shows that the average coronal X-ray luminosity decreases as the rotation period increases.
Just from this previous result of X-ray observations, we may possibly expect that superflare energy have a correlation with the stellar rotation
and rapidly rotating stars have more energetic flares.
However, Kepler data brought us new results somewhat different from this expectation, as summarized below.

\begin{figure}
	\centering
	\includegraphics[width=0.95\linewidth]{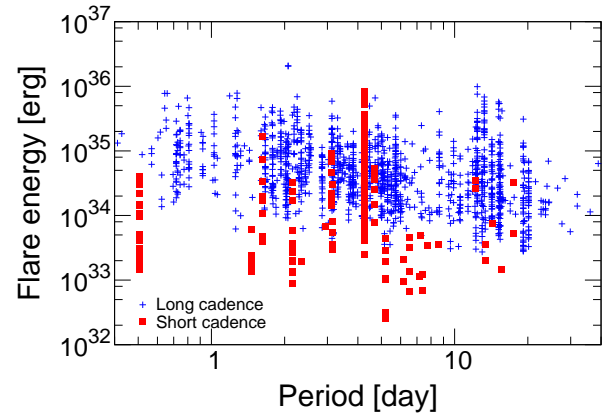}
	\caption{Scatter plot of the flare energy vs. the brightness variation period
(interpreted as the rotation period of each star).
Red filled-squares and blue small-crosses indicate superflares on solar-type stars
detected from short-cadence \citep{Maehara2015} and long-cadence data \citep{YNotsu2013}, respectively.
An apparent negative correlation between the variation period and the lower limit 
of the flare energy results from the detection limit of our flare search method.
}\label{fig:poster-fig9}
\end{figure}

\begin{figure*}
	\centering
	\includegraphics[width=0.48\linewidth]{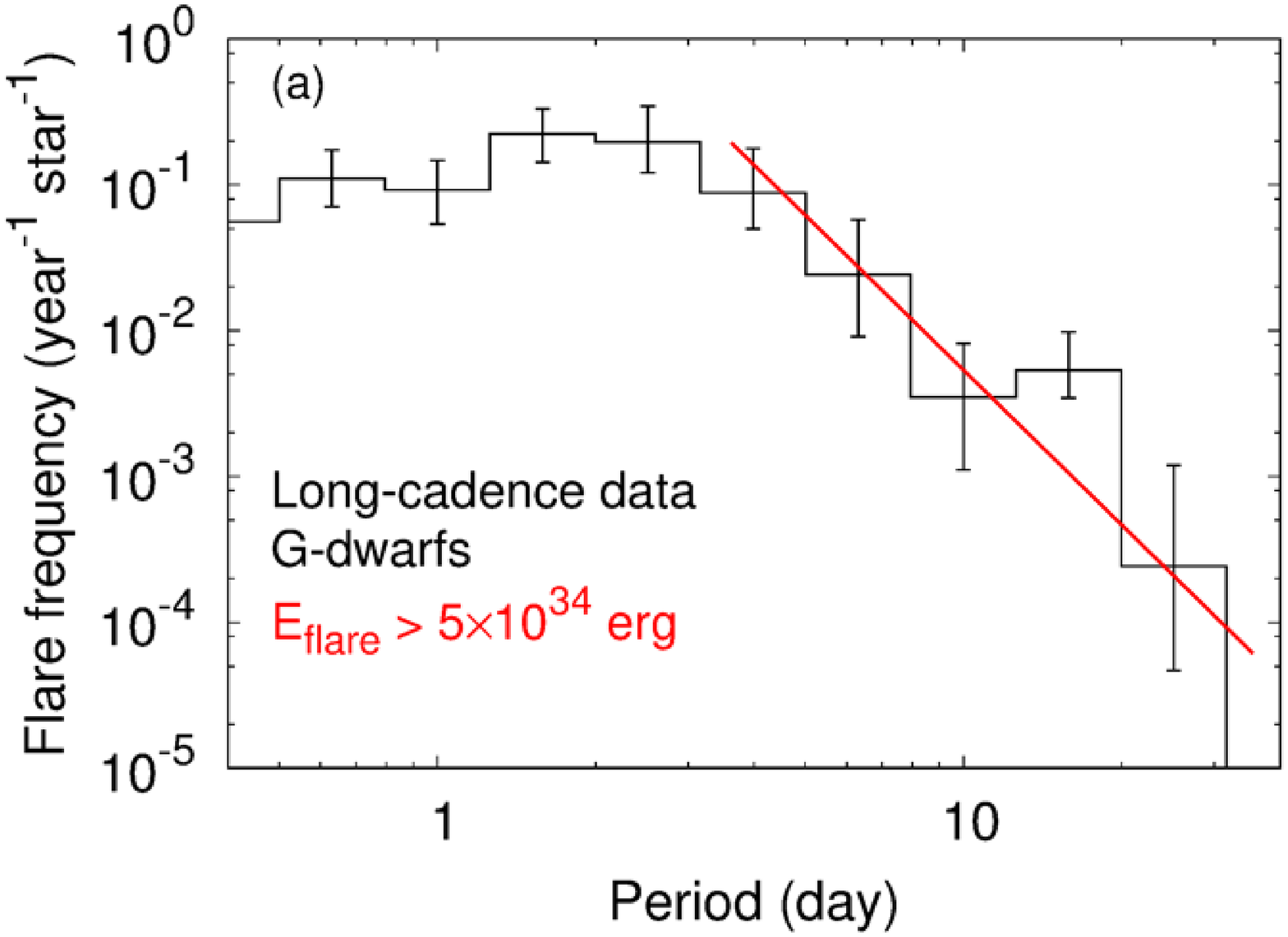}
	\includegraphics[width=0.48\linewidth]{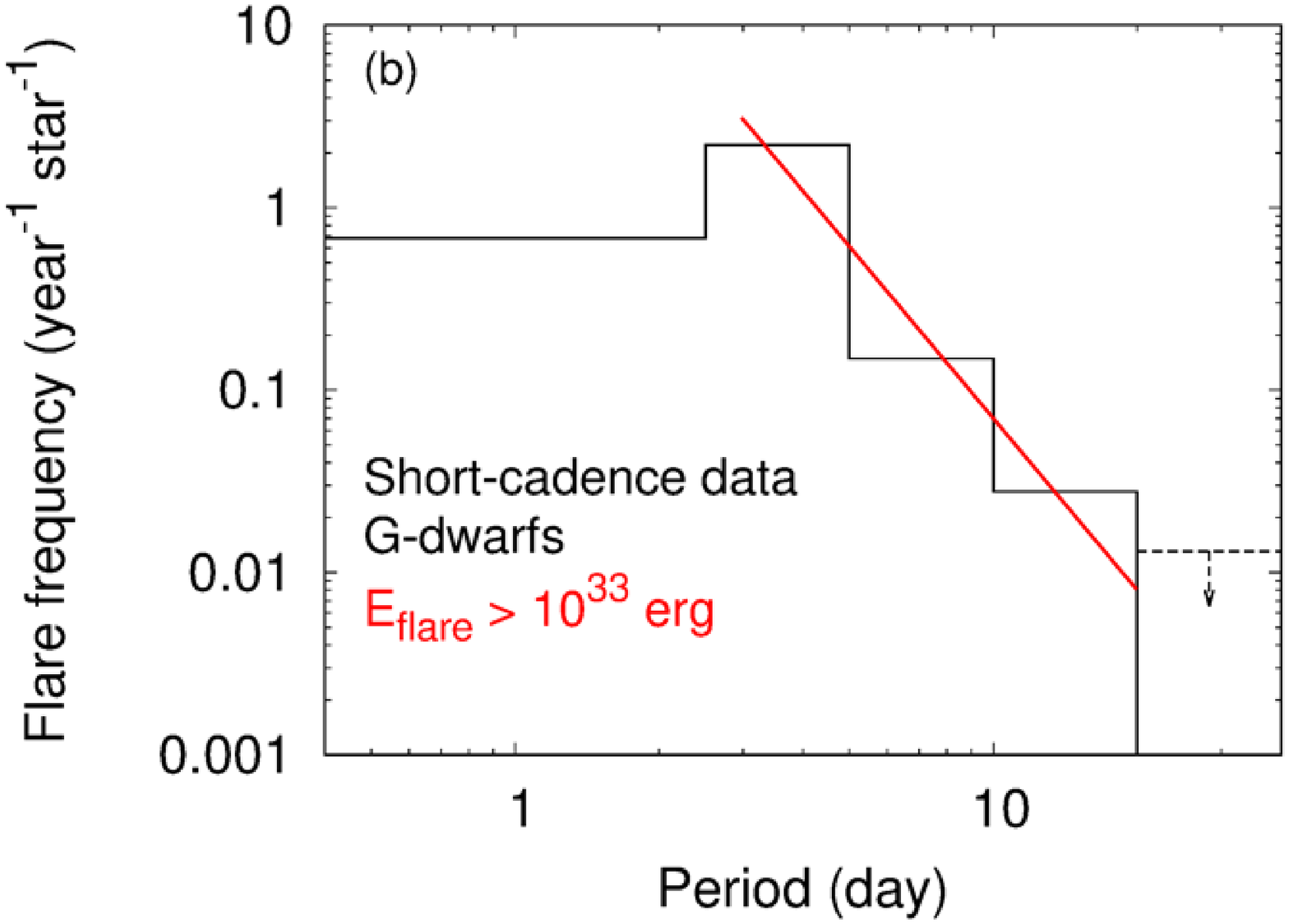}	
	\caption{(a) Distribution of the occurrence frequency of superflares as a function of 
the brightness variation period (rotational period of each star), using the data of superflares detected from Kepler long (30-min) cadence data \citep{YNotsu2013}.
The vertical axis indicates the number of flares with energy $>5\times10^{34}$ erg per star per year. 
The frequency distribution of superflares saturates for periods shorter than a few days. 
It is interesting to note that a similar saturation is observed for the relationship 
between the coronal X-ray activity and the rotation period of solar-type stars (cf. \citealt{Pizzolato2003}).
(b) Same as (a), but the plotted data are on the basis of the data of superflares 
detected from Kepler short (1-min) cadence data \citep{Maehara2015}.
The vertical axis indicates the number of flares with energy $>10^{33}$ erg per star per year. 
The dashed down arrow indicates the upper limit of the flare frequency.
}
	\label{fig:poster-fig7}
\end{figure*}

Figures \ref{fig:poster-fig9} \& \ref{fig:poster-fig7} show the relationship between the rotation period 
(estimated from the brightness variation) and the feature of superflares \citep{YNotsu2013}.
Figure \ref{fig:poster-fig9} indicates that the most energetic flare observed in a given rotation period bin
does not correlate with the stellar rotation period. 
At first glance, this is different from the expectation on the basis of X-ray observations (cf. \citealt{Pizzolato2003}) mentioned above.
If the superflare energy can be explained by the magnetic energy stored around the starspots (as shown in Figure \ref{fig:poster-fig8}),
this figure suggests that the maximum magnetic energy stored around the spot does not have a strong dependence on the rotation period.
Figure \ref{fig:poster-fig7}, however, shows that the average flare frequency in a given period bin 
tends to decrease as the period increases to longer than a few days. 
The frequency is averaged from all the superflare stars in the same period bin. 
The frequency of superflares on rapidly-rotating stars is higher than slowly-rotating stars,
and this indicates that a star evolves (and its rotational period increases), the frequency of superflares decreases
(cf. Similar tendency is also shown in \citet{Davenport2016}).
We can now interpret that this correlation between rotation period and flare frequency (not flare energy) 
is consistent with the correlation between rotation period and average X-ray luminosity seen in previous studies.
One amazing discovery from Figures \ref{fig:poster-fig9} \& \ref{fig:poster-fig7}  is that
superflares (with energy comparable to $10^{34}\sim10^{35}$ erg) can occur on stars rotating as slow as the Sun ($P_{\rm{rot}}=20\sim30$ days), 
even though the frequency is low (one in a few thousand years (cf. Figure \ref{fig:poster-fig6})), compared with rapidly-rotating stars.

\begin{table*}
	\centering
	\caption{The number of superflare stars having exoplanets (On the basis of the data retreived from the NASA Exoplanet Archive on May 29th, 2016).}
	\label{tab:flarestar-exoplanet}
	\begin{tabular*}{\linewidth}{c @{\extracolsep{\fill}} ccc}
	\noalign{\smallskip}\hline\hline\noalign{\smallskip}
	& All stars & \shortstack{Stars with \\ confirmed planets} & \shortstack{Stars with \\ candidate and confirmed planets} \\
	\noalign{\smallskip}\hline\noalign{\smallskip}
	\shortstack{All solar-type stars \\ ($T_{\rm{eff}}$=5300$\sim$6300K, $\log g\geq4.0$)\footnotemark} & \shortstack{102551 \\ {}} & \shortstack{1029 \\ (1.0\%)} & \shortstack{2131 \\ (2.1\%)} \\ \\
	\shortstack{Solar-type superflare stars \\ (\citealt{Shibayama2013})} & \shortstack{279 \\ {}} & \shortstack{1 (Kepler-491b) \\ (0.4\%)} & \shortstack{3 \\ (1.1\%)} \\
	\noalign{\smallskip}\hline
	\end{tabular*}
\end{table*}
	\footnotetext{
	In \citet{Shibayama2013}, we used stellar parameters taken from the original Kepler Input Catalog \citep{Brown2011},
	and the selection criteria of solar-type stars are $T_{\rm{eff}}$=5100$\sim$6000K, $\log g\geq4.0$. 
	However the temperatures in \citet{Brown2011} are systematically lower by 200 K than those in Revised Kepler Input Catalog \citep{Huber2014}. 
	In order to reduce the difference in statistical properties caused by the systematic difference in stellar temperature, 
	here we used the selection criteria of $T_{\rm{eff}}$=5300$\sim$6300K, $\log g\geq4.0$.}

\subsection{Superflare stars having exoplanets}\label{subsec:exoplantes}
\ \ \ \ \
\citet{Rubenstein2000} argued that solar-type stars 
with a hot Jupiter-like companion are good candidates for superflare stars; 
i.e., the hot Jupiter may play the role of a companion star in binary stars, such as in RS CVn stars,
which are magnetically very active, and produce many superflares.
\citet{Schaefer2000} concluded that the Sun has never produced superflares, 
since there is no hot Jupiter near to our Sun. 
We found, however, that fraction of superflare stars having planets (in particular, close-in giant planets like hot Jupiters)
are not higher than the fraction of ordinary solar-type stars having them, 
on the basis of Kepler data (Table \ref{tab:flarestar-exoplanet}).
This suggests hot Jupiter is not a necessary condition for superflares. 
In addition, \citet{Shibata2013} also found that hot Jupiters do not play any essential role in the generation of magnetic flux in the star itself, 
if we consider only the magnetic interaction between the star and the hot Jupiter, 
whereas the tidal interaction remains to be a possible cause of enhanced dynamo activity, though more detailed studies would be necessary.
This seems to be consistent with the above observational finding that hot Jupiter is not a necessary condition for superflares.

\section*{Acknowledgments}
{We are very grateful to the organizers of The Cool Stars 19 meeting. 
YN was supported by JSPS KAKENHI Grant Number JP16J00320 for attending this meeting.}

\end{document}